\documentstyle[times,epsfig]{jaa}
\begin{document}

\title[Unification of Radio Galaxies]{Unification of Radio Galaxies and Their Accretion/Jet Properties}
\author[Q. Wu, et al.]%
{Qingwen Wu$^{1}$\thanks{e-mail:qwwu@mail.hust.edu.cn}, Ya-Di Xu$^{2}$, \& Xinwu Cao$^{3}$ \\
        $^{1}$Physics Department, Huazhong University of Science and Technology, Wuhan, China\\
        $^{2}$Physics Department, Shanghai Jiao Tong University, Shanghai, China\\
        $^{3}$Shanghai Astronomical Observatory, CAS, Shanghai, China}

\maketitle
\label{firstpage}
\begin{abstract}
We investigate the relation between black hole mass, $M_{\rm bh}$, and jet power, $Q_{\rm jet}$, for a sample of
BL Lacs and radio quasars. We find that BL Lacs are separated from
radio quasars by the FR I/II dividing line in $M_{\rm bh}-Q_{\rm jet}$ plane, which strongly supports the
unification scheme of FR I/BL Lac and FR II/radio quasar. The Eddington ratio distribution
of BL Lacs and radio quasars exhibits a bimodal nature with a rough division at
$L_{\rm bol}/L_{\rm Edd}\sim 0.01$, which imply that they may have different accretion modes.
We calculate the jet power extracted from advection dominated accretion flow (ADAF),
and find that it require dimensionless angular momentum of black hole $j\simeq0.9-0.99$ to reproduce the dividing line
between FR I/II or BL  Lac/radio quasar if dimensionless accretion rate $\dot{m}=0.01$ is adopted, which is required by
above bimodal distribution of Eddington ratios. Our results suggest that black holes in radio galaxies
 are rapidly spinning.

\end{abstract}

\section{Introduction}
FR I radio galaxies (defined by edge-darkened radio structure) have
lower radio power than FR II galaxies (defined by edge-brightened
radio structure)(Fanaroff \& Riley 1974). In the frame of
unification schemes of radio-loud AGNs, low-power FR I radio galaxies
are believed to be misaligned BL Lacertae (BL Lac) objects, and high-power FR II
radio galaxies correspond to misaligned radio quasars (Urry \& Padovani 1995).
The unified scheme of radio galaxies have been extensively explored by many previous authors with different
approaches, such as, the comparisons of the spectral energy distributions in different wavebands
(e.g., Capetti et al. 2000), the radio morphology and luminosity functions (e.g., Padovani \& Urry 1991), etc.
Most BL Lacs have featureless optical/UV continuum spectra, and only
a small fraction of BL Lacs show very weak broad emission
lines, while quasars usually have strong broad line emission (e.g., V{\'e}ron-Cetty
\& V{\'e}ron 2000). The broad emission lines of quasars are produced by distant gas
 clouds in broad-line regions, which are photo-ionized by the optical/UV continua radiated
from the accretion disks. The difference of the broad line emission between radio-loud quasars and
BL Lacs may be attributed to their different central engines (e.g., Cao 2003).

FR I/II dichotomy is a famous observational result, which can be clearly divided in the host galaxy
optical luminosity and radio luminosity plane ( Ledlow \& Owen 1996). The physical reason for the FR I/II
division is still unclear, and there are two categories of models to explain it:
(1) their ambient medium  are different (e.g., Gopal-Krishna \& Wiita 2000); and/or (2) intrinsic difference of their central
engines, i.e., different accretion modes and/or jet formation processes (e.g., Meier 1999,
Ghisellini \& Celotti 2001).

\section{Unification of BL Lacs/Radio Quasars and FR I/II}

\begin{figure}
\hspace{3.0cm}
\epsfig{file=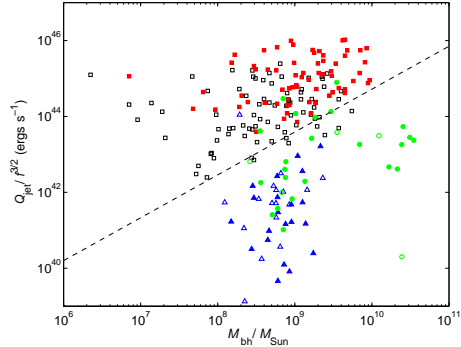,width=6.0cm}
\vspace{0.1cm}
\caption{The relation between $M_{\rm bh}$ and $Q_{\rm jet}$, for the BL Lacs and
quasars. The open squares and filled squares represent FSRQs and
SSRQs respectively, while the circles and triangles represent BL Lacs.
The filled circles/triangles represent the LBLs/HBLs with
measured line emission, while the open circles/triangles the
LBLs/HBLs without measured line emission. The dashed line represents
the Ledlow-Owen dividing line between FR I and FR II radio galaxies
in $M_{\rm bh}-Q_{\rm jet}$ plane. Parameter $f$ describe the uncertainties
 in deriving the jet power (see Wu \& Cao 2008).}
\label{fig1}
\end{figure}

 Our sample include 66 BL Lacs (28 low-energy-peaked BL Lacs, LBLs, and 38 high-energy-peaked BL Lacs, HBLs)
 and 146 radio quasars (79 flat-spectrum radio quasars, FSRQs, and 67 steep-spectrum radio quasars,
 SSRQs)(see Xu et al. 2009). Figure 1 show the relation between the black hole (BH)
 mass, $M_{\rm bh}$, and jet power, $Q_{\rm jet}$ for both BL Lacs and radio quasars.
 It is found that the BL Lacs can be well separated from radio quasars by the FR I/II dividing line.
 This result strongly supports the unification schemes for FR I/BL Lac and FR II/radio quasar.

\section{Bimodal Distribution of Eddington Ratios for Radio Galaxies}
The distribution of Eddington ratios for BL Lacs and radio quasars are shown
in Figure 2, where the bolometric luminosities of BL Lacs are derived from the
[O II] narrow line luminosities, while they are calculated from the broad line
luminosities for radio quasars, since that the continuum spectrum of radio
galaxies may be contaminated by the beaming emission from the relativistic jets (see
Xu et al. 2009). We find that the Eddington ratio distributions for BL Lacs
and quasars in our sample exhibit a bimodal nature. The BL Lacs are roughly separated
from the quasars at $L_{\rm bol}/L_{\rm Edd}\sim 0.01$, with most of BL Lac objects having
$L_{\rm bol}/L_{\rm Edd}\leq 0.01$, while almost all radio quasars having
$L_{\rm bol}/L_{\rm Edd}\geq 0.01$. We suggest that this bimodal behavior of the distribution
may imply different accretion modes in BL Lacs and radio quasars. The critical Eddington
ratio of $L_{\rm bol}/L_{\rm Edd}\sim 0.01$ is roughly consistent with theoretical prediction
for accretion mode transition from a standard accretion disk to an advection dominated accretion
flow (ADAF, e.g., Narayan \& Yi 1995). Therefore, our results suggest that ADAFs may present in
BL Lacs and standard thin disks are in radio quasars. A similar explanation is invoked
to explain the FR I/II dichotomy (e.g., Ghisellini \& Celotti 2001).

\begin{figure}
\hspace{3.0cm}
\epsfig{file=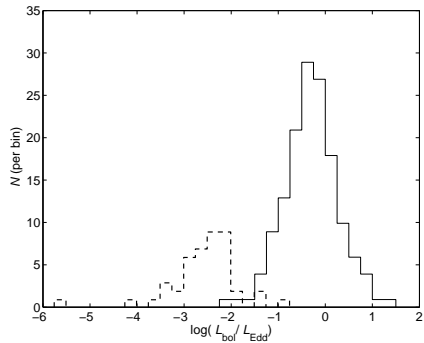,width=5.5cm}
\vspace{0.cm}
\caption{The $L_{\rm bol}/L_{\rm Edd}$ distributions for BL Lacs (dashed line)
and quasars (solid line).}
\label{fig1}
\end{figure}

\section{Jet Power Extracted from ADAF and FR I/II or BL Lac/Radio Quasar Dichotomy}
  We calculate the global structure of an accretion flow surrounding a spinning BH
  (see Wu \& Cao 2008 for more details, and references therein). Then we calculate
   the maximal jet power as a function of black hole mass with the hybrid jet formation
  model, which proposed by Meier (2001) as a variant of classical Blandford-Znajek
  model (BZ, Blandford \& Znajek 1977).  We find that it can roughly reproduce the dividing
  line of the FR I/II dichotomy (also the BL Lac and radio quasars as in Figure 1)
  in $Q_{\rm jet}-M_{\rm bh}$ plane, if the dimensionless accretion rate $\dot{m}\sim0.01$ and BH spin parameter
  $j\sim0.9-0.99$ are adopted (see Figure 3). We note that the $\dot{m}\sim0.01$
   is required to reproduce the bimodal distribution of Eddington ratios (Figure 2). Our result
   suggests that the BHs in radio galaxies are rapidly spinning.

\section{Acknowledgements}
 This work is supported by the National Basic Research Program of China (2009CB824800),
 the NSFC (grants 11143001, 10778621, 10703003, 11078014, 10773020, 10821302, and 10833002), the Science
  and Technology Commission of Shanghai Municipality (10XD1405000), the CAS (KJCX2-YW-T03) and HUST(0124-012030).

\begin{figure}
\hspace{3.0cm}
\epsfig{file=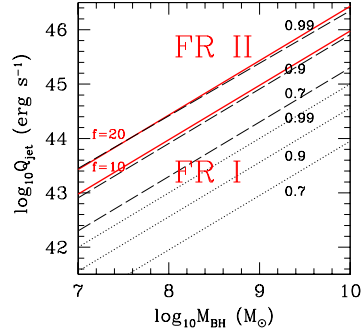,width=5.cm}
\vspace{0.cm}
\caption{ The jet power versus BH mass. The red solid lines represent the dividing line of
FR I/II or BL Lac/radio quasar, where the jet
power is calculated from 151 MHz radio luminosity with two different correction
 factors $f=20$ (upper) and $f=10$ (lower), respectively. The dashed and dotted
 lines are the jet power of the hybrid model and pure BZ model extracted from the
ADAFs with $j=0.99,0.9,0.7$ (from top to bottom), respectively.}
\label{fig1}
\end{figure}

\label{lastpage}
\end{document}